\documentclass[a4paper,twocolumn,11pt]{quantumarticle}
\pdfoutput=1
\usepackage[utf8]{inputenc}
\usepackage[english]{babel}
\usepackage[T1]{fontenc}
\usepackage{amsmath}
\usepackage{hyperref}
\usepackage{graphicx}
\usepackage{float} 
\usepackage{subfigure}
\usepackage{lipsum}
\usepackage{braket}

\begin{document}

\title{Variational Quantum Eigensolvers with Quantum Gaussian Filters for solving ground-state problems in quantum many-body systems}

\author{Yihao Liu}
\affiliation{Guangdong-Hong Kong Joint Laboratory of Quantum Matter, Department of Physics, and HK Institute of Quantum Science \&  Technology, The University of Hong Kong, Pokfulam Road, Hong Kong, China}
\orcid{0009-0006-3179-639X}
\author{Min-Quan He}
\email{hemq@hku.hk}
\orcid{0000-0002-0187-9024}
\affiliation{Guangdong-Hong Kong Joint Laboratory of Quantum Matter, Department of Physics, and HK Institute of Quantum Science \&  Technology, The University of Hong Kong, Pokfulam Road, Hong Kong, China}
\author{Z. D. Wang}
\email{zwang@hku.hk}
\affiliation{Guangdong-Hong Kong Joint Laboratory of Quantum Matter, Department of Physics, and HK Institute of Quantum Science \&  Technology, The University of Hong Kong, Pokfulam Road, Hong Kong, China}
\maketitle

\begin{abstract}
  We present a novel quantum algorithm for approximating the ground-state in quantum many-body systems, particularly suited for Noisy Intermediate-Scale Quantum~(NISQ) devices. Our approach integrates Variational Quantum Eigensolvers~(VQE) with Quantum Gaussian Filters~(QGF), utilizing an iterative methodology that discretizes the application of the QGF operator into small, optimized steps through VQE. Demonstrated on the Transverse Field Ising models, our method shows improved convergence speed and accuracy, particularly under noisy conditions, compared to conventional VQE methods. This advancement highlights the potential of our algorithm in effectively addressing complex quantum simulations, marking a significant stride in quantum computing applications within the NISQ era.
\end{abstract}

\section{Introduction}
\label{sec_introduction}

The eigenstate problem in quantum many-body systems stands as a key challenge with profound implications in fields ranging from quantum chemistry~\cite{mcardle2020quantum, aspuru2005simulated} to material science~\cite{bauer2020quantum} and the energy industry~\cite{kim2022fault, delgado2022simulating}. The inherent complexity of these problems, characterized by exponentially growing computational dimensions, has historically rendered exact solutions elusive. In this context, quantum computing, leveraging the principles of quantum mechanics, emerges as a promising avenue for more efficiently addressing these challenges compared to classical approaches.

Over the past decades, a plethora of quantum algorithms have been proposed to tackle the quantum many-body eigenstate problem. Notably, Quantum Phase Estimation~(QPE)~\cite{cleve1998quantum}, a cornerstone in this domain, offers the ability to accurately determine eigenenergy spectra and prepare corresponding eigenstates. However, the practical application of QPE is hindered by its substantial resource requirements, involving numerous controlled unitary operations and an inverse quantum Fourier transform, which exceed the capabilities of near-term quantum devices~\cite{huggins2021efficient, ma2020quantum}.

In response, various adaptations of QPE have been developed to mitigate resource demands~\cite{knill2007optimal, nagaj2009fast, poulin2009sampling, dorner2009optimal, abrams1999quantum}. Despite these advancements, the requirements remain formidable for current quantum technologies. This leads to the exploration of alternative approaches, particularly for problems centered around a specific eigenstate. Here, quantum filtering algorithms emerge as a more resource-efficient solution~\cite{zeng2021universal}. These algorithms operate by applying a non-unitary operator to an initial state, effectively filtering out undesired states while retaining the target state, especially the ground state. The construction of such non-unitary operators, derived from functions of the target Hamiltonian. Examples include imaginary time evolution~\cite{mcardle2019variational, motta2020determining, mao2023measurement, zhang2021continuous, tang2021qubit, parrish2019quantum, wang2019accelerated, zhang2022variational}, QGF algorithms~\cite{he2022quantum, apers2022quadratic, wang2022state, wang2023quantum}, cosine filter algorithms~\cite{ge2019faster, lu2021algorithms}, sine filter algorithms~\cite{xie2022power}, powered Hamiltonian operator algorithms~\cite{bespalova2021hamiltonian, seki2021quantum}, and quantum inverse iteration method~\cite{kyriienko2020quantum, he2022inverse}. The non-trivial implementation of these non-unitary operators on quantum computers has prompted methods like linear combination of unitaries~\cite{ge2019faster}, energy sequential estimation~\cite{kyriienko2020quantum}, and variational techniques~\cite{mcardle2019variational}.

In the era of NISQ devices~\cite{preskill2018quantum}, where circuit depth is constrained by the presence of noise, VQE~\cite{peruzzo2014variational, kandala2017hardware, tilly2022variational, zhang2020collective} have gained prominence. VQE employs a shallow circuit depth, utilizing the variational principle to optimize the parameters of a parametrized state, thereby solving the ground-state problem. The VQE framework, comprising a trial state evolved through a parametrized circuit, also known as ansatz, and classical optimization of parameters to minimize a cost function, aligns well with the implementation of quantum filtering operators in the NISQ era.

This paper introduces a novel method for preparing the ground state of many-body systems using a shallow circuit. Our approach discretizes the QGF operator into a series of small-step evolutions. Each step involves using VQE to simulate the evolution process, updating the parameters so that the quantum state approximates the evolved state after each small step. Starting from an initial state represented by initial parameters, the iterative updating of parameters enables the approximation of the QGF-evolved state, ultimately converging to the ground state of the target Hamiltonian.

The paper is structured as follows: Section \ref{sec_quantum_algorithm} details the principles and procedures of our quantum algorithm. The algorithm's efficacy in solving the ground-state problem is demonstrated numerically in Section \ref{sec_numerical_demonstration}. Finally, Section \ref{sec_conclusion} concludes the paper and discusses future implications.

\section{Quantum Algorithm}
\label{sec_quantum_algorithm} 

In this section, we introduce a novel quantum algorithm leveraging the QGF method to address ground-state problems in many-body quantum systems. Our approach is centered around the principle of QGF, its application in converging an initial quantum state towards a target ground state, and the discrete implementation of this operator through a series of stepwise evolutions. We further elucidate the integration of VQE in simulating these step evolutions and outline the comprehensive procedure of our proposed algorithm.

\subsection{QGF algorithms}
\label{sub_qgf_algorithms}

The QGF method is conceptualized to solve eigenstate challenges in many-body systems, where the Hamiltonian \(\hat{H}\) is represented as \(\hat{H} = \sum_{l} c_{l} \hat{h}_{l}\), with each \(\hat{h}_{l}\) denoting a Pauli string. The essence of this method lies in its ability to selectively filter eigenstates of \(\hat{H}\) using a Gaussian function of the Hamiltonian, formulated as \(e^{-\hat{H}^{2}/\sigma^{2}}\). When this quantum gaussian filter operator is applied to an eigenstate \(\ket{\lambda_{j}}\) of \(\hat{H}\), with \(\lambda_{j}\) being its eigenvalue, the resultant state becomes \(e^{-\lambda_{j}^{2}/\sigma^{2}}\ket{\lambda_{j}}\). This approach effectively re-weights the amplitude of each eigenstate in a given initial state, represented as a superposition \(\ket{\phi_{i}} = \sum_{j=0}^{2^{n}-1} a_{j} \ket{\lambda_{j}}\).

Through this re-weighting, the QGF method enables the preparation of a desired eigenstate. The resulting state, post-application of the filter, is described as 
\begin{equation}
    \label{eq:qgf_result}
    \ket{\phi_{f}} = \frac{1}{\sqrt{C}} \sum_{j=0}^{2^{n}-1} a_{j} e^{-\lambda_{j}^{2}/\sigma^{2}} \ket{\lambda_{j}}.
\end{equation}
Here, the factor \(e^{-\lambda_{j}^{2}/\sigma^{2}}\) serves as the additional weight induced by the quantum Gaussian filter, while \(\frac{1}{\sqrt{C}} = (\sum_{j=0}^{2^{n}-1} a_{j} e^{-\lambda_{j}^{2}/\sigma^{2}})^{\frac{1}{2}}\) maintains the state's normalization. Notably, the Gaussian function applied to the eigenvalues shows a monotonically decreasing trend for values greater than zero. Therefore, in cases where all eigenvalues are positive, eigenstates with smaller eigenvalues receive greater emphasis. As the magnitude of the eigenvalues or the parameter \(\sigma^{2}\) increases, the resultant state increasingly resembles the ground state. This characteristic positions the QGF as a potent tool for addressing ground-state problems.

A crucial aspect to consider is the non-unitary nature of the QGF operator, which presents a challenge for direct implementation on quantum computers. To address this, we propose an iterative approximation method, which we discuss in subsequent sections.

Given the non-unitary nature of the QGF, direct application on quantum computers remains challenging. To circumvent this, we adopt an iterative method for approximate implementation. The Gaussian filter operator, traditionally represented as \(e^{-\hat{H}^{2}/\sigma^{2}}\), is reformulated to \(e^{-\hat{H}^{2} \tau}\), where \(\tau = 1/\sigma^{2}\) represents the inverse of the Gaussian function's variance. The evolved state under this operator is expressed as \(\ket{\phi(\tau)} = \frac{1}{\sqrt{C}} e^{-\hat{H}^{2} \tau} \ket{\phi(0)}\), with \(\ket{\phi(0)}\) denoting the initial state and \(\frac{1}{\sqrt{C}}\) ensuring normalization.

Instead of evolving the state directly from \(0\) to \(\tau\), we decompose this process into \(k = \tau/\Delta\tau\) smaller steps, each involving an incremental evolution of the state by \(e^{-\hat{H}^{2} \Delta\tau}\). The differential evolution of the state with respect to \(\tau\) is given by \(\frac{\partial \ket{\phi(\tau)}}{\partial \tau} = -\hat{H}^{2} \ket{\phi(\tau)}\). For a small incremental change in \(\tau\) by \(\Delta \tau\), the state's evolution can be approximated as \(\ket{\phi(\tau + \Delta \tau)} \approx \frac{1}{\sqrt{C}} (\ket{\phi(\tau)} - \Delta \tau \hat{H}^{2} \ket{\phi(\tau)})\), where the normalization factor \(\frac{1}{\sqrt{C}}\) is recalculated at each step.

\subsection{VQE Implementation for Step Evolution}
\label{sub_vqe_implementation_for_step_evolution}
The VQE framework involves utilizing a parameterized quantum circuit to simulate the evolution of an initial state, with the classical process optimizing the circuit parameters against a predefined cost function. The quantum state is parameterized as \(\ket{\psi(\vec{\theta})} = \hat{U}(\vec{\theta})\ket{\vec{0}}\), where \(\hat{U}(\vec{\theta})\) represents the unitary operations dependent on the parameter vector \(\vec{\theta}\). A commonly used cost function in VQE is the estimated energy, \(E(\vec{\theta}) = \braket{\psi(\vec{\theta})|\hat{H}|\psi(\vec{\theta})}\), with the objective of finding the parameter set that minimizes this energy.

In our approach, the initial state is defined as \(\ket{\phi(0)} = \ket{\psi(\vec{\theta}^{(0)})}\). The goal is to iteratively evolve this state through \(T\) steps, from \(\ket{\phi(0)}\) to \(\ket{\phi(T\Delta\tau)}\). At each step, we aim to adjust the parameters such that the new state maximizes its overlap with the theoretically evolved state. For the \(i\)-th step, the prepared state \(\ket{\psi(\vec{\theta}^{(t)})}\) represents \(\ket{\phi(t\Delta\tau)}\). The objective is to adjust the parameters from \(\vec{\theta}^{(t)}\) to \(\vec{\theta}^{(t+1)}\) such that \(\ket{\phi((t+1)\Delta\tau)}\) closely matches the approximated evolved state \(\frac{1}{\sqrt{C}} (\ket{\phi(t\Delta\tau)} - \Delta \tau \hat{H}^{2} \ket{\phi(t\Delta\tau)})\).

This leads to a tailored cost function for training the VQE at each evolutionary step, aiming to maximize the overlap with the approximated evolved state. The cost function is thus formulated as:
\begin{equation}
    \label{eq:cost_function}
    \begin{split}
    C(\vec{\theta}^{(t+1)}) =
    &- |\braket{\psi(\vec{\theta}^{(t+1)})|\psi(\vec{\theta}^{(t)})}\\
    &- \Delta \tau \braket{\psi(\vec{\theta}^{(t+1)})|\hat{H}^{2}|\psi(\vec{\theta}^{(t)})}|.
    \end{split}
\end{equation}

Classical optimization techniques such as gradient descent and the ADAM optimizer~\cite{kingma2014adam} are utilized in the training process, with the Hadamard test~\cite{aharonov2006polynomial} employed for estimating components of the cost function. The optimization aims to find a parameter set that minimizes this cost function at each iteration, gradually refining the state towards the desired ground state.

\subsection{Incorporating McLachlan's Variational Principle in VQE}
\label{sub:mclachlan_principle}

In addition to the previously described VQE approach, we employ McLachlan's variational principle~\cite{mcardle2019variational, mclachlan1964variational, broeckhove1988equivalence} for a more nuanced evolution of the system's parameters. This principle provides an alternative and sophisticated method for parameter optimization, especially in the context of quantum state evolution.

McLachlan's variational principle asserts that the optimal parameter adjustments can be calculated using the equation \(\dot{\vec{\theta}} = A^{-1} \vec{C}\). Here, \(A\) is a matrix representing the overlap of the differential states, and \(\vec{C}\) is a vector capturing the influence of the Hamiltonian's square on the state evolution.

The matrix \(A\) is defined as:
\begin{equation}
    A_{j, k} = \Re\left(\frac{\partial \bra{\psi(\vec{\theta}^{(t)})}}{\partial \theta_{j}} \frac{\partial \ket{\psi(\vec{\theta}^{(t)})}}{\partial \theta_{k}}\right),
\end{equation}
and can be calculated using finite difference stochastic approximation~\cite{kiefer1952stochastic, finck2012performance} or analytical gradient measurements~\cite{romero2018strategies, li2017efficient}. The elements of \(A\) are essential for understanding the correlation between different parameter adjustments.

The computation of \(\vec{C}\) is similarly intricate, defined as:
\begin{equation}
    C_{j} = \Re\left(-\sum_{l,m} c_{l} c_{m} \frac{\partial \bra{\psi(\vec{\theta}^{(t)})}}{\partial \theta_{j}} \hat{h}_{l}\hat{h}_{m} \ket{\psi(\vec{\theta}^{(t)})}\right).
\end{equation}
This computation involves evaluating the impact of the squared Hamiltonian on the differential states, with the Hamiltonian decomposed into a sum of Pauli string coefficients.

The differential states required for computing both \(A\) and \(\vec{C}\) are prepared by strategically manipulating the variational circuit, particularly through the insertion of Pauli operators at specific locations within the circuit. This allows for the precise measurement of the impact of each parameter on the state evolution.

Once \(A\) and \(\vec{C}\) are computed, we obtain the gradient of the parameters, \(\dot{\vec{\theta}}\), which guides the update of the parameters for the next step. The parameter update formula is given by \(\vec{\theta}^{(t+1)} = \vec{\theta}^{(t)} + \Delta \tau \dot{\vec{\theta}}^{(t)}\). This process enables a meticulously controlled evolution of the quantum state, ensuring that it closely follows the desired trajectory as dictated by the quantum Gaussian filter method.

\subsection{Procedure of the Quantum Algorithm}
\label{sub_procedure_of_the_quantum_algorithm}

The implementation of our proposed quantum algorithm is a structured process designed to effectively approximate the ground state of a target many-body system. This subsection delineates the sequence of steps integral to the algorithm's execution.

\textbf{Parameter Determination}: Initially, the algorithm requires the specification of key parameters. These include the evolution range \(\tau\), the step size \(\Delta \tau\) for the quantum Gaussian filter evolution, and the selection of an appropriate VQE ansatz. This selection is influenced by both the model of the system under study and the quantum hardware available. Additionally, the method for optimizing the VQE, such as gradient descent, ADAM optimizer, or McLachlan’s variational principle, is also determined at this stage.

\textbf{Initial State Preparation}: The process commences with the preparation of an initial quantum state \(\ket{\psi_{i}}\). This state is tailored to the system's specifications and is generated by applying a suitable unitary operation \(\hat{U}_{i}\) to the default quantum state \(\ket{\vec{0}}\). An initial set of variational parameters \(\vec{\theta^{(0)}}\) is also chosen to represent the starting point of the state evolution.

\textbf{Iterative State Evolution}: The core of the algorithm is the iterative evolution of the quantum state. At each step, the current trial state \(\ket{\phi(\vec{\theta}^{(t)})}\) representing \(\ket{\phi(t\Delta\tau)}\) undergoes optimization to align closely with the target state \(\ket{\phi((t+1)\Delta\tau)}\). This optimization process is governed by the VQE framework, aiming to maximize the overlap with the evolved state and adjust the variational parameters accordingly. 

Throughout this iterative process, the algorithm employs the defined cost function and optimization technique to refine the trial state. Each iteration involves adjusting the parameters based on the gradient computed either through conventional methods or McLachlan's variational principle, ensuring that the state evolution is in line with the trajectory established by the QGF method.

\textbf{Final State Approximation}: Upon completing the specified number of iterations, the algorithm yields a trial state \(\ket{\psi(\vec{\theta}^{T})}\) that approximates the ground state of the target system. This final state is the culmination of successive refinements through the iterative process and embodies the algorithm's ability to converge to the desired ground state.

\section{Numerical Demonstration}
\label{sec_numerical_demonstration}

\begin{figure*}
    \centering
    \subfigure[]{
    \begin{minipage}[b]{0.48\textwidth}
        \includegraphics[width=1\textwidth]{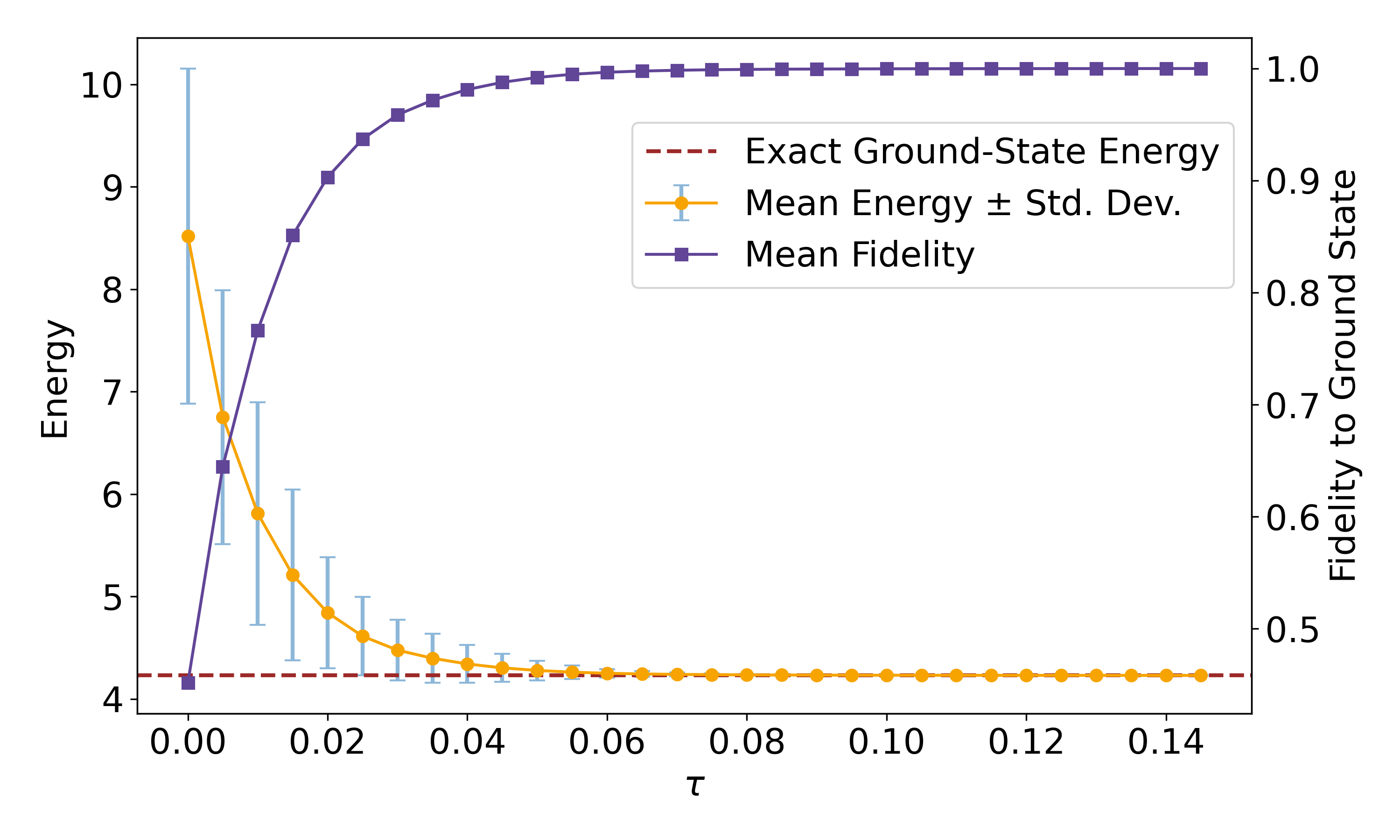}
        \label{fig:demo_ising_n4_hx}
    \end{minipage}}
    \subfigure[]{
    \begin{minipage}[b]{0.48\textwidth}
        \includegraphics[width=1\textwidth]{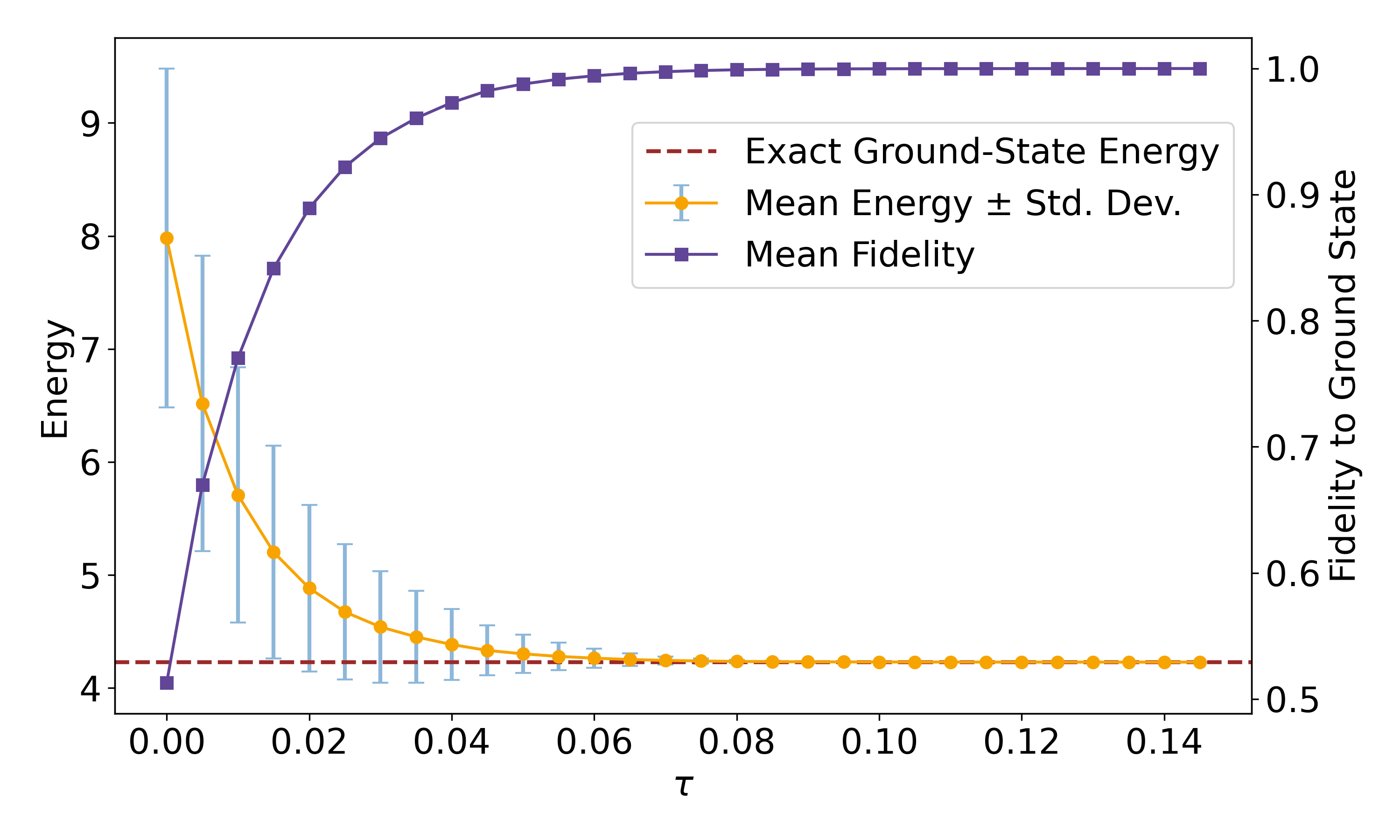}
        \label{fig:demo_ising_n4_hzz}
    \end{minipage}}
    \subfigure[]{
    \begin{minipage}[b]{0.48\textwidth}
        \includegraphics[width=1\textwidth]{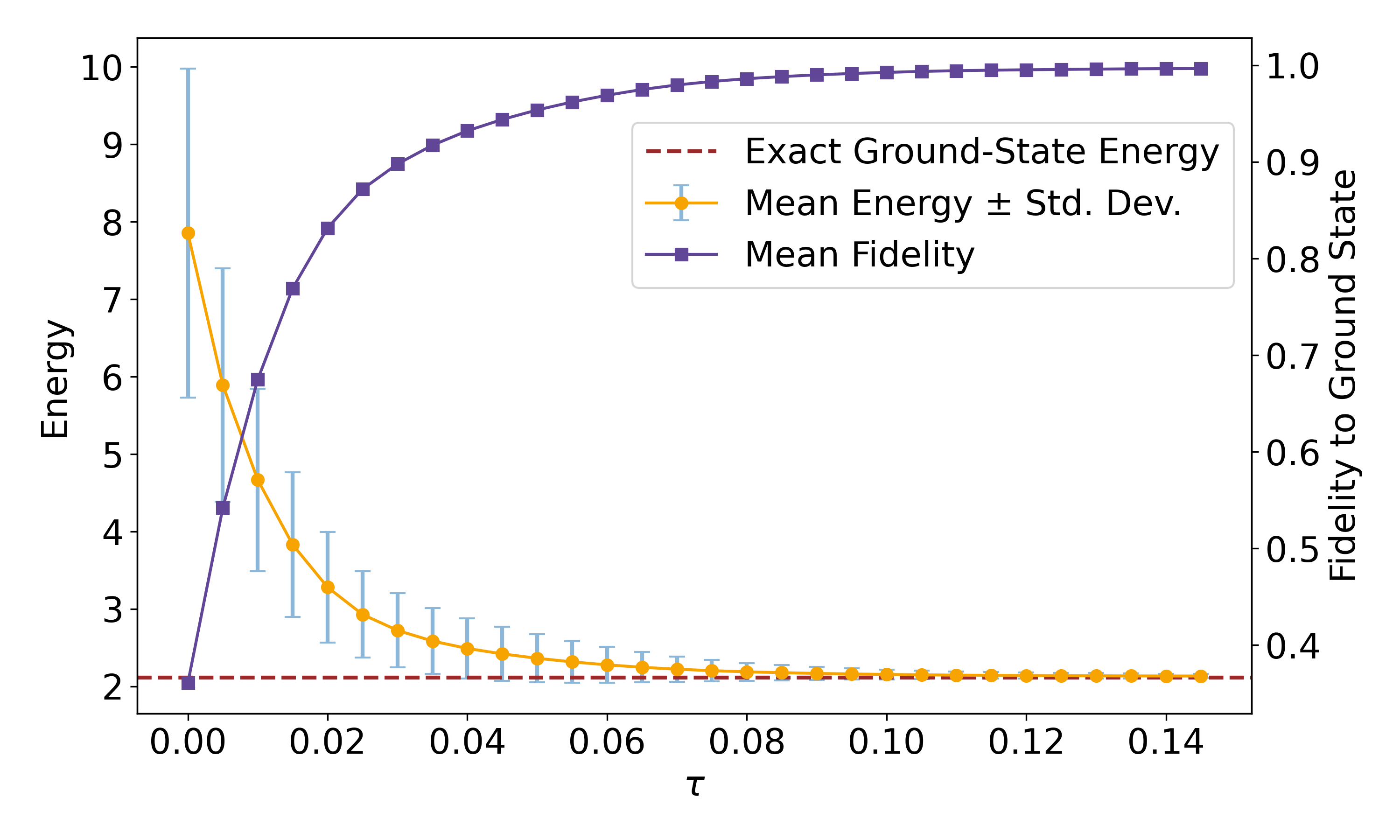}
        \label{fig:demo_ising_n6_hx}
    \end{minipage}}
    \subfigure[]{
    \begin{minipage}[b]{0.48\textwidth}
        \includegraphics[width=1\textwidth]{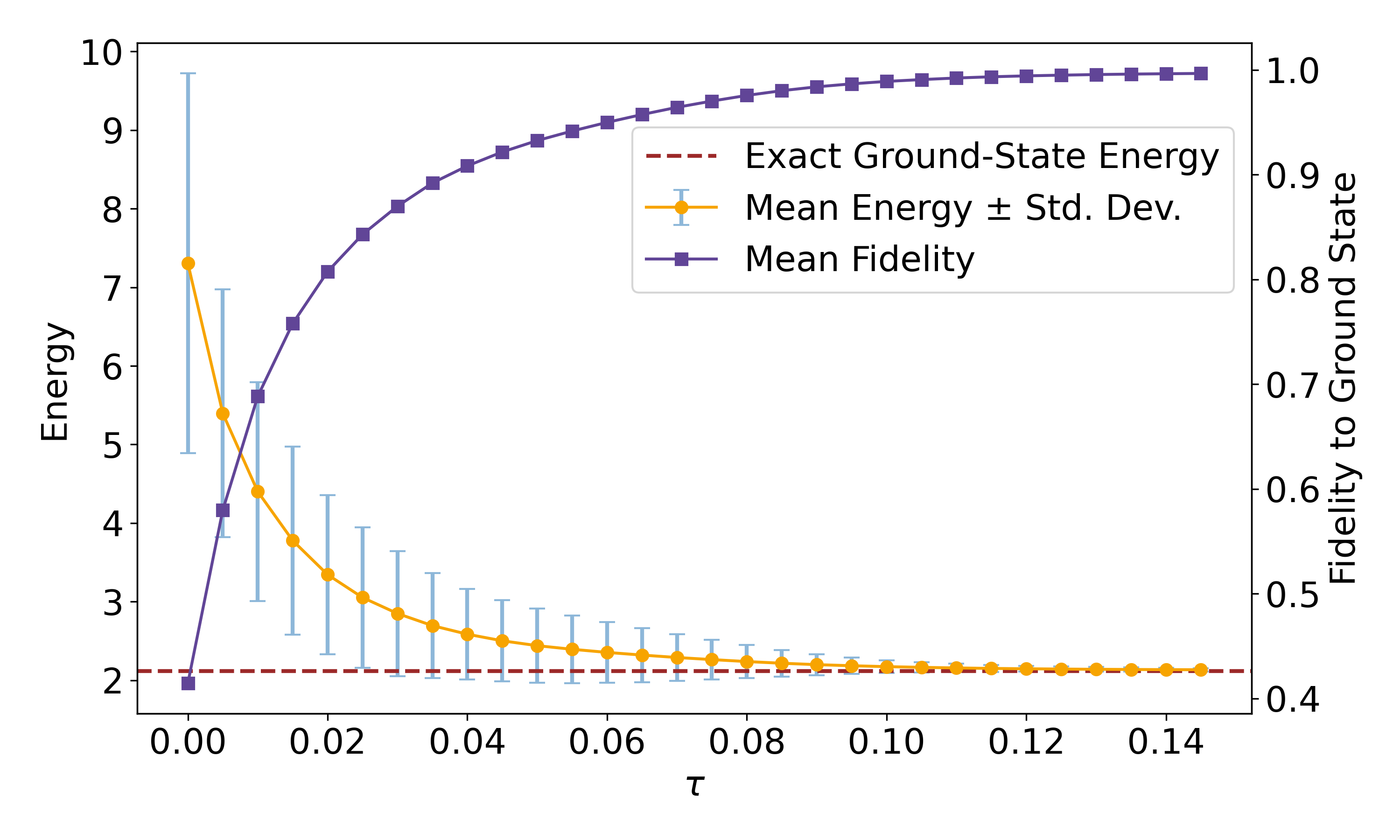}
        \label{fig:demo_ising_n6_hzz}
    \end{minipage}}
    
    \caption{Numerical results for the Transverse Field Ising Model. The yellow line represents the mean value of the estimated ground-state energy. The blue error bars indicate the standard deviation of the estimated energy. The red dashed line marks the exact ground-state energy, providing a reference for comparison. The purple line depicts the mean fidelity between the prepared state and the exact ground state. (a) illustrates the model with $N = 4$ qubits in the ferromagnetic phase, (b) shows the model with $N = 4$ qubits in the paramagnetic phase, (c) illustrates the model with $N = 6$ qubits in the ferromagnetic phase, while and (d) shows the model with $N = 6$ qubits in the paramagnetic phase.}
    \label{fig:demo_ising}
\end{figure*}

In this section, we provide a numerical demonstration of our algorithm designed for solving the ground state in quantum many-body models. Our focus is primarily on the Transverse Field Ising Model. The simulations are conducted using the open-sourced quantum simulation package QuTiP~\cite{johansson2013qutip}. The Hamiltonian of the Transverse Field Ising Model~\cite{ising1924beitrag} is represented as:
\begin{equation}
    \label{eq:ising_model_hamiltonian}
    \hat{H} = -J \sum_{n = 1}^{N} \hat{\sigma}_{n}^{z} \hat{\sigma}_{n + 1}^{z} + g \sum_{n = 1}^{N} \hat{\sigma}_{n}^{x},
\end{equation}
where $J$ is the interaction strength of nearby sites, $g$ is the scale of the external transverse field, and $\hat{\sigma}^{z}_{N+1} = \hat{\sigma}^{z}_{1}$ is for periodic boundary condition. 
Two sets of parameters are investigated: $J = 0.5, g = 1$, which leads to a dominant $g$ term indicative of the paramagnetic phase, and $J = 1, g = 0.5$, where the $J$ term prevails, signifying the ferromagnetic phase.

We select system sizes of $N = 4$ and $N = 6$. For both cases, we apply an energy shift of $8.5$ to move the ground-state energy to approximately $4.228$ for the $4$-qubit case and $2.115$ for the $6$-qubit case. The initial state for each phase is prepared with specific considerations. For the ferromagnetic phase, characterized by spin alignment, each qubit is initialized in the ground state of $\hat{\sigma}^{x}$ by applying a Pauli-Z gate after a Hadamard gate to $\ket{0}$, i.e. $\ket{\psi_{i}} = \otimes_{n=1}^{N} \hat{\sigma}_{n}^{z} H \ket{0}$. In the paramagnetic phase, where spins align more with the external field, we create a GHZ state using Hadamard and CNOT gates, followed by a phase-flip gate on each qubit. A $p=4$ layer QAOA ansatz~\cite{hadfield2019quantum} is employed. The parameterized state is described as $\ket{\psi(\vec{\theta})} = [\prod_{p=0}^{3} e^{-i \theta_{2p} \hat{h}_{x}} e^{-i \theta_{2p+1} \hat{h}_{zz}}] \ket{\psi_{i}}$, where $\hat{h}_{zz}$ and $\hat{h}_{x}$ correspond to the first and second terms of Eq.~\eqref{eq:ising_model_hamiltonian}, respectively, and $\ket{\psi_{i}}$ is the prepared initial state.

The initial parameters $\vec{\theta^{(0)}}$ are generated randomly, with $50$ distinct sets for each simulation. We evolve the quantum Gaussian filter for $\tau = 0.15$ with a step size of $\Delta \tau = 0.005$, resulting in a total of $30$ evolution steps. At each step, we minimize the cost function Eq.~\eqref{eq:cost_function} using gradient descent, limiting iterations to a maximum of $10$ per step. If there is no improvement in cost, the training process is interrupted, and we proceed to the next evolution step.

Fig.~\ref{fig:demo_ising} displays the results of our algorithm. It shows the mean and standard deviation of the estimated ground-state energy at each evolution step. Additionally, it presents the fidelity between the algorithmically prepared state and the exact ground state. These results affirm that our method effectively evolves different initial states towards the target ground state, and can handle the scale effect as well.

\begin{figure}
    \centering
    \subfigure[]{
    \begin{minipage}[b]{0.5\textwidth}
        \includegraphics[width=1\textwidth]{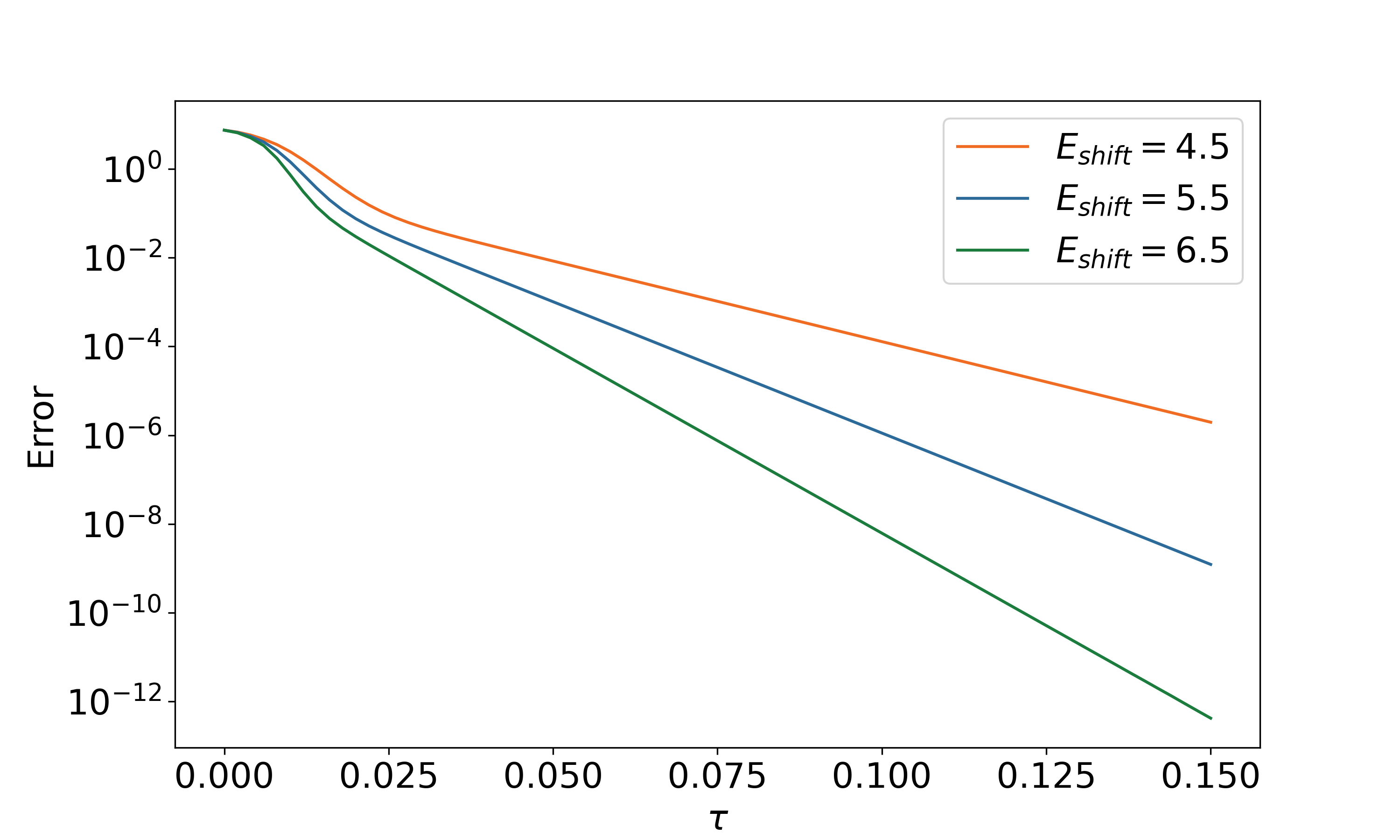}
        \label{fig:effect_shift_Hx}
    \end{minipage}}
    \subfigure[]{
    \begin{minipage}[b]{0.5\textwidth}
        \includegraphics[width=1\textwidth]{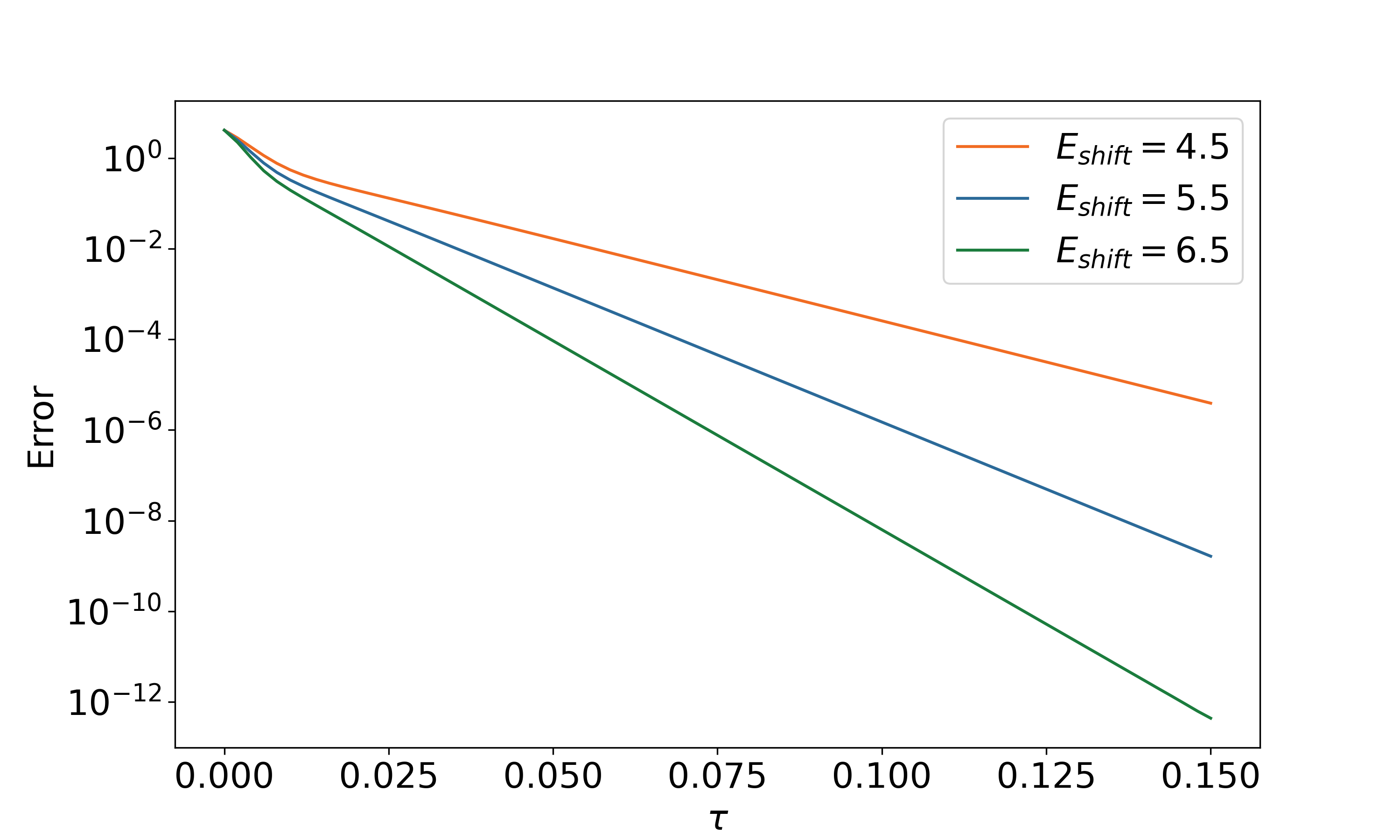}
        \label{fig:effect_shift_Hzz}
    \end{minipage}}
    
    \caption{
        This figure illustrates the numerical results for the Transverse Field Ising Model with $N=4$, focusing on the influence of different shift-energy values on the algorithm's performance. The depicted lines represent the energy error as a function of the evolution parameter $\tau$, with each line corresponding to a distinct shift-energy: orange for $4.5$, blue for $5.5$, and green for $6.5$. The results are split into two parts: (a) demonstrates the performance in the ferromagnetic phase, while (b) displays the outcomes in the paramagnetic phase. The trend observed across these graphs confirms that an increase in shift-energy correlates with a more rapid convergence towards the desired state, validating the theoretical predictions associated with our algorithm.
        }
    \label{fig:effect_shift}
\end{figure}

In our investigation, we also analyze the impact of varying shift-energy values. According to Eq.~\eqref{eq:qgf_result}, the effective weighting of various eigenstates, including the ground state, is influenced by the energy scale. An increase in shift-energy generally reduces the weight of additional eigenstates while accentuating the relative importance of the ground state. This adjustment can potentially enhance the performance of the algorithm. However, a higher shift-energy also necessitates approximating the QGF filter over a broader energy range, requiring a smaller step size $\Delta\tau$. In this demonstration, we employ the $N = 4$ qubit model with shift-energy values of $4.5$, $5.5$, and $6.5$, aiming to adjust the ground-state energy around $0.228$, $1.228$, and $2.228$, respectively. We set $\Delta\tau = 0.002$ and evolve the QGF to $\tau = 0.15$. For parameter updates, we utilize the McLachlan method. Fig.~\ref{fig:effect_shift} presents our numerical results, showcasing the estimated energy error as a function of evolution $\tau$. These findings indicate that a higher shift-energy leads to a faster convergence rate, aligning with our theoretical expectations.

\begin{figure}
    \centering
    \subfigure[]{
    \begin{minipage}[b]{0.5\textwidth}
        \includegraphics[width=1\textwidth]{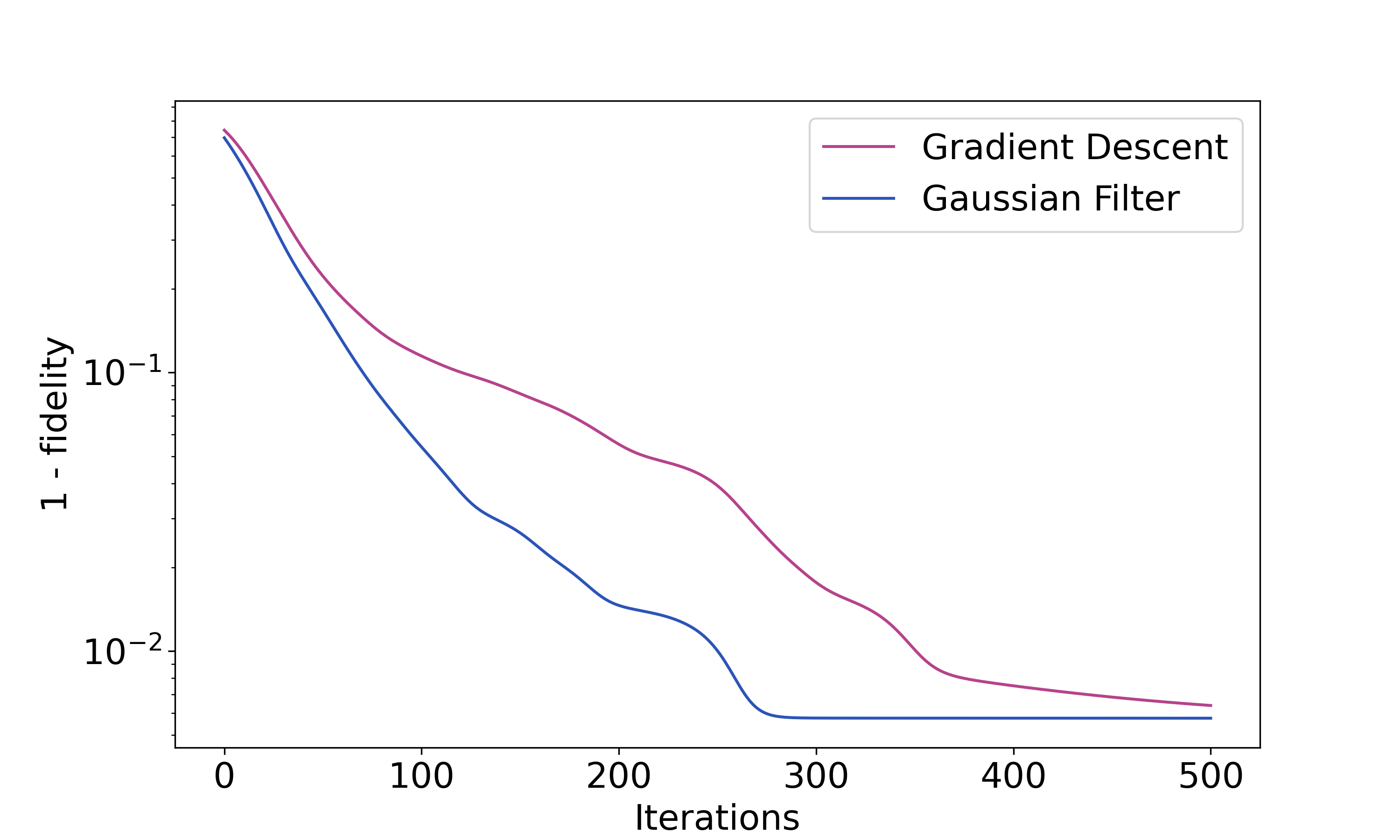}
        \label{fig:simulation_noisy_random}
    \end{minipage}}
    \subfigure[]{
    \begin{minipage}[b]{0.5\textwidth}
        \includegraphics[width=1\textwidth]{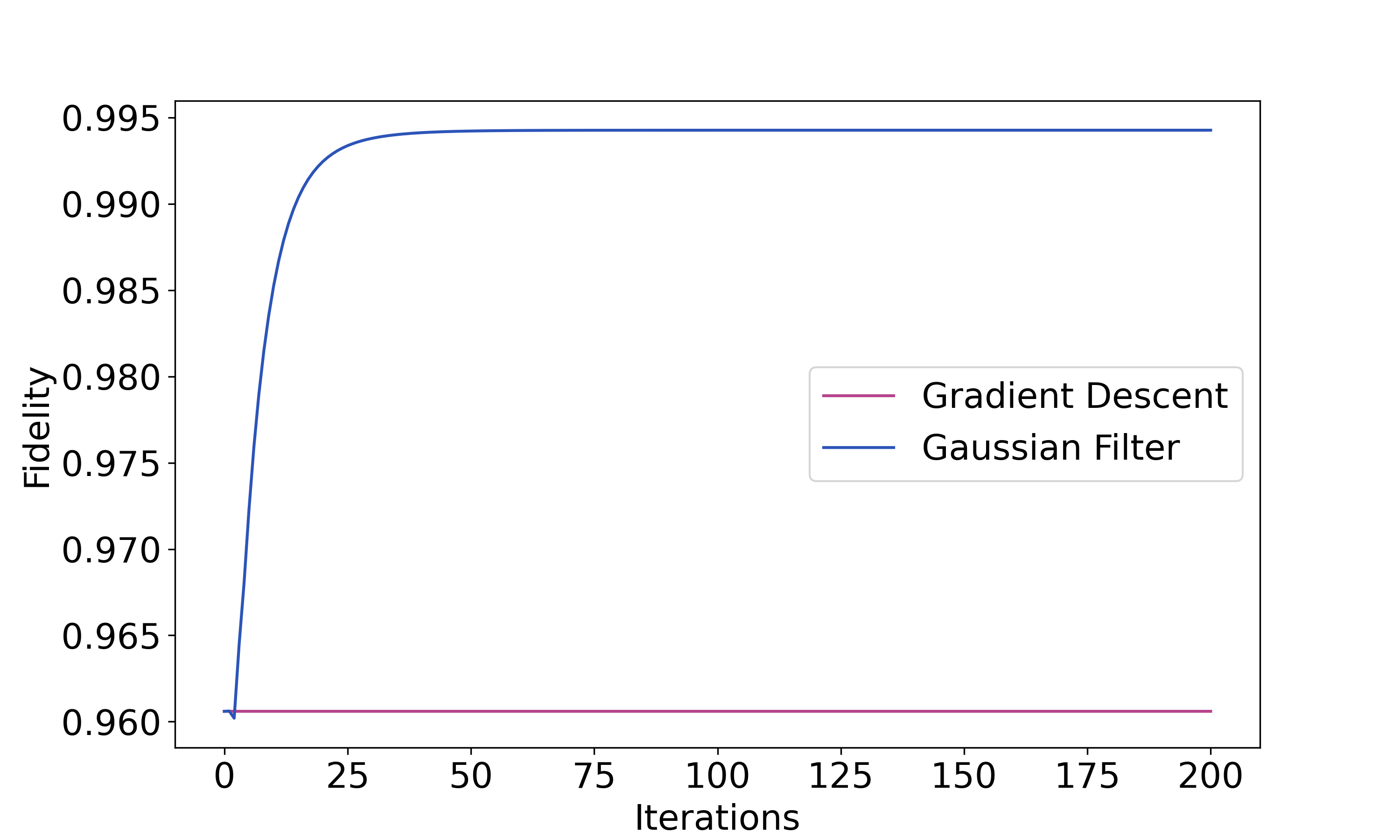}
        \label{fig:simulation_noisy_random_prepared}
    \end{minipage}}
    
    \caption{Numerical simulation results for the Transverse Field Ising Model with a $N=4$, conducted in a depolarizing noise environment characterized by a noise probability, $p_{l} = 0.0001$, after each state rotation. This figure contrasts the performance of the conventional gradient descent method (represented by the pink line) with our proposed algorithm (blue line). The results are split into two parts: (a) illustrates outcomes using $100$ random initial parameter seeds, depicting the average fidelity between the prepared state and the exact ground state, while (b) presents results from initial parameters set to zero with minor perturbations, representing a well-prepared initial state, and shows the fidelity between the prepared state and the exact ground state.}
    \label{fig:simulation_noisy}
\end{figure}

We extend our study to simulate the algorithm's efficacy in preparing the ground state of the Transverse Field Ising Model in the paramagnetic phase under a noisy environment. The parameters for the Hamiltonian are set as $N = 4$, $g = 1$, $J = 0.5$, with a shift-energy $E_{s} = 11$. We employ a $4$-layer QAOA ansatz for this simulation. During the state rotation process, a depolarization noise characterized by a probability $p_{l} = 0.0001$ is introduced. The evolution step size is set at $\Delta \tau = 0.005$, with the total number of iterations capped at 500.

To evaluate the robustness and efficiency of our proposed method, we compare it with a standard VQE implementation using gradient descent with learning rate $l = 0.01$ and $500$ iteration steps. For both methods, $100$ random initial parameter sets are sampled. Most of these parameter sets converge to stable points, with only a few converging to suboptimal results characterized by significantly lower fidelity to the exact ground state. These outlier results can be filtered out based on their abnormally high estimated energy values.

The comparative results are illustrated in Fig.~\ref{fig:simulation_noisy_random}, where the average fidelity between the exact ground state and the prepared state is plotted. The orange line represents the results from the gradient descent method, while the blue line corresponds to our proposed algorithm. Notably, our method demonstrates superior performance compared to the traditional VQE approach, both in terms of convergence speed and the quality of the converged result. The achieved fidelity in our method is approximately $0.995$, indicating a high degree of accuracy in ground state preparation under noisy conditions.

Furthermore, we investigated the efficiency of ground state preparation using both methods when initiated from a well-prepared state, characterized by initial parameters set to zero with minor perturbations. As the result shown in Fig.~\ref{fig:simulation_noisy_random_prepared}, we observed that the standard method tends to quickly become trapped in a local minimum, characterized by the gradient approximating a zero vector, which significantly impedes the optimization process. In contrast, our method consistently avoids this pitfall, effectively optimizing the parameters from the outset. This observation suggests that our algorithm, when initiated from an appropriately chosen initial state, requires fewer optimization steps to achieve convergence, thereby enhancing its efficiency in practical applications.

\section{Conclusion}
\label{sec_conclusion}

This research presents a novel quantum algorithm designed for solving the ground-state problem of quantum many-body systems, particularly tailored for NISQ devices. The core of our approach is the VQE method, enhanced through the implementation of the QGF operator. This operator's unique capability to effectively converge an initial quantum state towards the ground state of a target Hamiltonian underpins the algorithm's efficacy.

Our method's distinctiveness lies in its iterative process, discretizing the QGF operator's application into a series of small evolutionary steps. Each step employs the VQE framework, leveraging a parameterized quantum circuit and classical optimization techniques to iteratively refine the quantum state. This process progressively aligns the state with the theoretical trajectory established by the QGF method, culminating in a high-fidelity approximation of the ground state.

Numerical demonstrations on the Transverse Field Ising Model substantiate the algorithm's potential. Our simulations, including scenarios with depolarizing noise, exhibit promising results. The algorithm consistently produces states with high fidelity to the exact ground states, demonstrating its robustness and adaptability to various quantum many-body models and noisy environments. Compared to trivial VQE method, our approach has a faster converge rate and a higher expected fidelity. Notably, it performs exceptionally well when initiated from a well-prepared state, suggesting broader applicability to other scenarios with similarly prepared initial states, as discussed in Ref.~\cite{tubman2018postponing, babbush2015chemical, sugisaki2018quantum, mcardle2020quantum}.

In conclusion, our quantum algorithm stands as a significant advancement in the pursuit of practical quantum computing solutions for complex quantum many-body problems. Its compatibility with NISQ devices, coupled with its demonstrated effectiveness, positions it as a valuable tool for future explorations in quantum computation reliant on accurate quantum state simulations. Additionally, our algorithm has the versatility to prepare other eigenstates by introducing a shift in energy near the target state.

\acknowledgements
We thank Dan-Bo Zhang, Yan Zhu, and Yu-Cheng Chen for valuable discussions. This work was supported by the Guangdong-Hong Kong Joint Laboratory of Quantum Matter, the NSFC/RGC JRS grant~(RGC Grant No. N\_HKU774/21, NSFC Grant No. 12161160315), the  CRF~(Grant No. C6009-20G) and GRF~(Grant No. 17310622) of Hong Kong.

\bibliographystyle{unsrt}
\bibliography{reference.bib}

\begin{thebibliography}{10}

\bibitem{mcardle2020quantum}
Sam McArdle, Suguru Endo, Al{\'a}n Aspuru-Guzik, Simon~C Benjamin, and Xiao
  Yuan.
\newblock Quantum computational chemistry.
\newblock {\em Reviews of Modern Physics}, 92(1):015003, 2020.

\bibitem{aspuru2005simulated}
Al{\'a}n Aspuru-Guzik, Anthony~D Dutoi, Peter~J Love, and Martin Head-Gordon.
\newblock Simulated quantum computation of molecular energies.
\newblock {\em Science}, 309(5741):1704--1707, 2005.

\bibitem{bauer2020quantum}
Bela Bauer, Sergey Bravyi, Mario Motta, and Garnet Kin-Lic Chan.
\newblock Quantum algorithms for quantum chemistry and quantum materials
  science.
\newblock {\em Chemical Reviews}, 120(22):12685--12717, 2020.

\bibitem{kim2022fault}
Isaac~H Kim, Ye-Hua Liu, Sam Pallister, William Pol, Sam Roberts, and Eunseok
  Lee.
\newblock Fault-tolerant resource estimate for quantum chemical simulations:
  Case study on li-ion battery electrolyte molecules.
\newblock {\em Physical Review Research}, 4(2):023019, 2022.

\bibitem{delgado2022simulating}
Alain Delgado, Pablo~AM Casares, Roberto Dos~Reis, Modjtaba~Shokrian Zini,
  Roberto Campos, Norge Cruz-Hern{\'a}ndez, Arne-Christian Voigt, Angus Lowe,
  Soran Jahangiri, Miguel~Angel Martin-Delgado, et~al.
\newblock Simulating key properties of lithium-ion batteries with a
  fault-tolerant quantum computer.
\newblock {\em Physical Review A}, 106(3):032428, 2022.

\bibitem{cleve1998quantum}
Richard Cleve, Artur Ekert, Chiara Macchiavello, and Michele Mosca.
\newblock Quantum algorithms revisited.
\newblock {\em Proceedings of the Royal Society of London. Series A:
  Mathematical, Physical and Engineering Sciences}, 454(1969):339--354, 1998.

\bibitem{huggins2021efficient}
William~J Huggins, Jarrod~R McClean, Nicholas~C Rubin, Zhang Jiang, Nathan
  Wiebe, K~Birgitta Whaley, and Ryan Babbush.
\newblock Efficient and noise resilient measurements for quantum chemistry on
  near-term quantum computers.
\newblock {\em npj Quantum Information}, 7(1):23, 2021.

\bibitem{ma2020quantum}
He~Ma, Marco Govoni, and Giulia Galli.
\newblock Quantum simulations of materials on near-term quantum computers.
\newblock {\em npj Computational Materials}, 6(1):85, 2020.

\bibitem{knill2007optimal}
Emanuel Knill, Gerardo Ortiz, and Rolando~D Somma.
\newblock Optimal quantum measurements of expectation values of observables.
\newblock {\em Physical Review A}, 75(1):012328, 2007.

\bibitem{nagaj2009fast}
Daniel Nagaj, Pawel Wocjan, and Yong Zhang.
\newblock Fast amplification of qma.
\newblock {\em arXiv preprint arXiv:0904.1549}, 2009.

\bibitem{poulin2009sampling}
David Poulin and Pawel Wocjan.
\newblock Sampling from the thermal quantum gibbs state and evaluating
  partition functions with a quantum computer.
\newblock {\em Physical review letters}, 103(22):220502, 2009.

\bibitem{dorner2009optimal}
Uwe Dorner, Rafal Demkowicz-Dobrzanski, Brian~J Smith, Jeff~S Lundeen, Wojciech
  Wasilewski, Konrad Banaszek, and Ian~A Walmsley.
\newblock Optimal quantum phase estimation.
\newblock {\em Physical review letters}, 102(4):040403, 2009.

\bibitem{abrams1999quantum}
Daniel~S Abrams and Seth Lloyd.
\newblock Quantum algorithm providing exponential speed increase for finding
  eigenvalues and eigenvectors.
\newblock {\em Physical Review Letters}, 83(24):5162, 1999.

\bibitem{zeng2021universal}
Pei Zeng, Jinzhao Sun, and Xiao Yuan.
\newblock Universal quantum algorithmic cooling on a quantum computer.
\newblock {\em arXiv preprint arXiv:2109.15304}, 2021.

\bibitem{mcardle2019variational}
Sam McArdle, Tyson Jones, Suguru Endo, Ying Li, Simon~C Benjamin, and Xiao
  Yuan.
\newblock Variational ansatz-based quantum simulation of imaginary time
  evolution.
\newblock {\em npj Quantum Information}, 5(1):75, 2019.

\bibitem{motta2020determining}
Mario Motta, Chong Sun, Adrian~TK Tan, Matthew~J O’Rourke, Erika Ye, Austin~J
  Minnich, Fernando~GSL Brandao, and Garnet Kin-Lic Chan.
\newblock Determining eigenstates and thermal states on a quantum computer
  using quantum imaginary time evolution.
\newblock {\em Nature Physics}, 16(2):205--210, 2020.

\bibitem{mao2023measurement}
Yuping Mao, Manish Chaudhary, Manikandan Kondappan, Junheng Shi, Ebubechukwu~O
  Ilo-Okeke, Valentin Ivannikov, and Tim Byrnes.
\newblock Measurement-based deterministic imaginary time evolution.
\newblock {\em Physical Review Letters}, 131(11):110602, 2023.

\bibitem{zhang2021continuous}
Dan-Bo Zhang, Guo-Qing Zhang, Zheng-Yuan Xue, Shi-Liang Zhu, and ZD~Wang.
\newblock Continuous-variable assisted thermal quantum simulation.
\newblock {\em Physical review letters}, 127(2):020502, 2021.

\bibitem{tang2021qubit}
Ho~Lun Tang, VO~Shkolnikov, George~S Barron, Harper~R Grimsley, Nicholas~J
  Mayhall, Edwin Barnes, and Sophia~E Economou.
\newblock qubit-adapt-vqe: An adaptive algorithm for constructing
  hardware-efficient ans{\"a}tze on a quantum processor.
\newblock {\em PRX Quantum}, 2(2):020310, 2021.

\bibitem{parrish2019quantum}
Robert~M Parrish, Edward~G Hohenstein, Peter~L McMahon, and Todd~J
  Mart{\'\i}nez.
\newblock Quantum computation of electronic transitions using a variational
  quantum eigensolver.
\newblock {\em Physical review letters}, 122(23):230401, 2019.

\bibitem{wang2019accelerated}
Daochen Wang, Oscar Higgott, and Stephen Brierley.
\newblock Accelerated variational quantum eigensolver.
\newblock {\em Physical review letters}, 122(14):140504, 2019.

\bibitem{zhang2022variational}
Yu~Zhang, Lukasz Cincio, Christian~FA Negre, Piotr Czarnik, Patrick~J Coles,
  Petr~M Anisimov, Susan~M Mniszewski, Sergei Tretiak, and Pavel~A Dub.
\newblock Variational quantum eigensolver with reduced circuit complexity.
\newblock {\em npj Quantum Information}, 8(1):96, 2022.

\bibitem{he2022quantum}
Min-Quan He, Dan-Bo Zhang, and Z.~D. Wang.
\newblock Quantum gaussian filter for exploring ground-state properties.
\newblock {\em Phys. Rev. A}, 106:032420, Sep 2022.

\bibitem{apers2022quadratic}
Simon Apers, Shantanav Chakraborty, Leonardo Novo, and J{\'e}r{\'e}mie Roland.
\newblock Quadratic speedup for spatial search by continuous-time quantum walk.
\newblock {\em Physical review letters}, 129(16):160502, 2022.

\bibitem{wang2022state}
Guoming Wang, Sukin Sim, and Peter~D Johnson.
\newblock State preparation boosters for early fault-tolerant quantum
  computation.
\newblock {\em Quantum}, 6:829, 2022.

\bibitem{wang2023quantum}
Guoming Wang, Daniel~Stilck Fran{\c{c}}a, Ruizhe Zhang, Shuchen Zhu, and
  Peter~D Johnson.
\newblock Quantum algorithm for ground state energy estimation using circuit
  depth with exponentially improved dependence on precision.
\newblock {\em Quantum}, 7:1167, 2023.

\bibitem{ge2019faster}
Yimin Ge, Jordi Tura, and J~Ignacio Cirac.
\newblock Faster ground state preparation and high-precision ground energy
  estimation with fewer qubits.
\newblock {\em Journal of Mathematical Physics}, 60(2), 2019.

\bibitem{lu2021algorithms}
Sirui Lu, Mari~Carmen Banuls, and J~Ignacio Cirac.
\newblock Algorithms for quantum simulation at finite energies.
\newblock {\em PRX Quantum}, 2(2):020321, 2021.

\bibitem{xie2022power}
Qing-Xing Xie, Yi~Song, and Yan Zhao.
\newblock Power of the sine hamiltonian operator for estimating the eigenstate
  energies on quantum computers.
\newblock {\em Journal of Chemical Theory and Computation}, 18(12):7586--7602,
  2022.

\bibitem{bespalova2021hamiltonian}
Tatiana~A Bespalova and Oleksandr Kyriienko.
\newblock Hamiltonian operator approximation for energy measurement and
  ground-state preparation.
\newblock {\em PRX Quantum}, 2(3):030318, 2021.

\bibitem{seki2021quantum}
Kazuhiro Seki and Seiji Yunoki.
\newblock Quantum power method by a superposition of time-evolved states.
\newblock {\em PRX Quantum}, 2(1):010333, 2021.

\bibitem{kyriienko2020quantum}
Oleksandr Kyriienko.
\newblock Quantum inverse iteration algorithm for programmable quantum
  simulators.
\newblock {\em npj Quantum Information}, 6(1):7, 2020.

\bibitem{he2022inverse}
Min-Quan He, Dan-Bo Zhang, and ZD~Wang.
\newblock Inverse iteration quantum eigensolvers assisted with a continuous
  variable.
\newblock {\em Quantum Science and Technology}, 7(2):025026, 2022.

\bibitem{preskill2018quantum}
John Preskill.
\newblock Quantum computing in the nisq era and beyond.
\newblock {\em Quantum}, 2:79, 2018.

\bibitem{peruzzo2014variational}
Alberto Peruzzo, Jarrod McClean, Peter Shadbolt, Man-Hong Yung, Xiao-Qi Zhou,
  Peter~J Love, Al{\'a}n Aspuru-Guzik, and Jeremy~L O’brien.
\newblock A variational eigenvalue solver on a photonic quantum processor.
\newblock {\em Nature communications}, 5(1):4213, 2014.

\bibitem{kandala2017hardware}
Abhinav Kandala, Antonio Mezzacapo, Kristan Temme, Maika Takita, Markus Brink,
  Jerry~M Chow, and Jay~M Gambetta.
\newblock Hardware-efficient variational quantum eigensolver for small
  molecules and quantum magnets.
\newblock {\em nature}, 549(7671):242--246, 2017.

\bibitem{tilly2022variational}
Jules Tilly, Hongxiang Chen, Shuxiang Cao, Dario Picozzi, Kanav Setia, Ying Li,
  Edward Grant, Leonard Wossnig, Ivan Rungger, George~H Booth, et~al.
\newblock The variational quantum eigensolver: a review of methods and best
  practices.
\newblock {\em Physics Reports}, 986:1--128, 2022.

\bibitem{zhang2020collective}
Dan-Bo Zhang and Tao Yin.
\newblock Collective optimization for variational quantum eigensolvers.
\newblock {\em Physical Review A}, 101(3):032311, 2020.

\bibitem{kingma2014adam}
Diederik~P Kingma and Jimmy Ba.
\newblock Adam: A method for stochastic optimization.
\newblock {\em arXiv preprint arXiv:1412.6980}, 2014.

\bibitem{aharonov2006polynomial}
Dorit Aharonov, Vaughan Jones, and Zeph Landau.
\newblock A polynomial quantum algorithm for approximating the jones
  polynomial.
\newblock In {\em Proceedings of the thirty-eighth annual ACM symposium on
  Theory of computing}, pages 427--436, 2006.

\bibitem{mclachlan1964variational}
AD~McLachlan.
\newblock A variational solution of the time-dependent schrodinger equation.
\newblock {\em Molecular Physics}, 8(1):39--44, 1964.

\bibitem{broeckhove1988equivalence}
J~Broeckhove, L~Lathouwers, E~Kesteloot, and P~Van~Leuven.
\newblock On the equivalence of time-dependent variational principles.
\newblock {\em Chemical physics letters}, 149(5-6):547--550, 1988.

\bibitem{kiefer1952stochastic}
Jack Kiefer and Jacob Wolfowitz.
\newblock Stochastic estimation of the maximum of a regression function.
\newblock {\em The Annals of Mathematical Statistics}, pages 462--466, 1952.

\bibitem{finck2012performance}
Steffen Finck and Hans-Georg Beyer.
\newblock Performance analysis of the simultaneous perturbation stochastic
  approximation algorithm on the noisy sphere model.
\newblock {\em Theoretical computer science}, 419:50--72, 2012.

\bibitem{romero2018strategies}
Jonathan Romero, Ryan Babbush, Jarrod~R McClean, Cornelius Hempel, Peter~J
  Love, and Al{\'a}n Aspuru-Guzik.
\newblock Strategies for quantum computing molecular energies using the unitary
  coupled cluster ansatz.
\newblock {\em Quantum Science and Technology}, 4(1):014008, 2018.

\bibitem{li2017efficient}
Ying Li and Simon~C Benjamin.
\newblock Efficient variational quantum simulator incorporating active error
  minimization.
\newblock {\em Physical Review X}, 7(2):021050, 2017.

\bibitem{johansson2013qutip}
J.R. Johansson, P.D. Nation, and Franco Nori.
\newblock Qutip 2: A python framework for the dynamics of open quantum systems.
\newblock {\em Comput. Phys. Commun.}, 184(4):1234--1240, 2013.

\bibitem{ising1924beitrag}
Ernst Ising.
\newblock {\em Beitrag zur theorie des ferro-und paramagnetismus}.
\newblock PhD thesis, Grefe \& Tiedemann Hamburg, Germany, 1924.

\bibitem{hadfield2019quantum}
Stuart Hadfield, Zhihui Wang, Bryan O’gorman, Eleanor~G Rieffel, Davide
  Venturelli, and Rupak Biswas.
\newblock From the quantum approximate optimization algorithm to a quantum
  alternating operator ansatz.
\newblock {\em Algorithms}, 12(2):34, 2019.

\bibitem{tubman2018postponing}
Norm~M Tubman, Carlos Mejuto-Zaera, Jeffrey~M Epstein, Diptarka Hait, Daniel~S
  Levine, William Huggins, Zhang Jiang, Jarrod~R McClean, Ryan Babbush, Martin
  Head-Gordon, et~al.
\newblock Postponing the orthogonality catastrophe: efficient state preparation
  for electronic structure simulations on quantum devices.
\newblock {\em arXiv preprint arXiv:1809.05523}, 2018.

\bibitem{babbush2015chemical}
Ryan Babbush, Jarrod McClean, Dave Wecker, Al{\'a}n Aspuru-Guzik, and Nathan
  Wiebe.
\newblock Chemical basis of trotter-suzuki errors in quantum chemistry
  simulation.
\newblock {\em Physical Review A}, 91(2):022311, 2015.

\bibitem{sugisaki2018quantum}
Kenji Sugisaki, Shigeaki Nakazawa, Kazuo Toyota, Kazunobu Sato, Daisuke Shiomi,
  and Takeji Takui.
\newblock Quantum chemistry on quantum computers: A method for preparation of
  multiconfigurational wave functions on quantum computers without performing
  post-hartree--fock calculations.
\newblock {\em ACS central science}, 5(1):167--175, 2018.

\end{thebibliography}
\end{document}